\pdfoutput=1 
\documentclass[12pt]{article}

\usepackage[english]{babel}
\usepackage{hyperref}
\usepackage{xcolor}
\usepackage{pslatex}
\usepackage{amsmath}
\usepackage{amssymb}
\usepackage{bbm}
\usepackage{euscript}
\usepackage{epsfig}
\usepackage{graphicx}
\usepackage{setspace}
\usepackage{cite}
\newcommand{\bhlumi}{{\tt BHLUMI}}
\newcommand{\bhwide}{{\tt BHWIDE}}
\newcommand{\babayaga}{{\tt BabaYaga}}

\newcommand{\order}[1]{${\cal O}(#1)$}

\begin{document}
\begin{titlepage}

%\begin{center}
%{\bf\LARGE ***** DRAFT \today\ ***** }
%\end{center}
%
\begin{flushright}
\bf IFJPAN-IV-2018-07, BU-HEPP-18-03, MPP-2018-91
\end{flushright}

% ---------------- Title and authors ----------------
\vspace{3mm}
\begin{center}
    {\Large\bf 
       The Path to $0.01\%$ Theoretical Luminosity Precision\\
       for the FCC-ee}$^{\star}$ 
\end{center}

% ---------------- Authors ----------------
\vskip 7mm
\begin{center}
{\large S.\ Jadach$^a$,
        W.\ P\l{}aczek$^b$,
        M.\ Skrzypek$^a$,
        B.F.L.\ Ward$^{c,d}$ and 
        S.A.\ Yost,$^e$ }
\\
\vskip 2mm
{\em $^a$Institute of Nuclear Physics, Polish Academy of Sciences,\\
  ul.\ Radzikowskiego 152, 31-342 Krak\'ow, Poland}
\\
\vspace{1mm}
{\em $^b$Marian Smoluchowski Institute of Physics, Jagiellonian University,\\
ul.\ \L{}ojasiewicza 11, 30-348 Krak\'ow, Poland}
\\
\vspace{1mm}
{\em $^c$Baylor University, Waco, TX, USA}
\\
\vspace{1mm}
{\em $^d$Max Planck Institute f\"ur Physik, M\"unchen, Germany}
\\
\vspace{1mm}
{\em $^e$The Citadel, Charleston, SC, USA}
\end{center}
 
\vspace{25mm}
\begin{abstract}
\noindent
The current status of the theoretical precision for the Bhabha
luminometry is critically reviewed and pathways are outlined
to the requirement targeted by the FCC-ee precision studies.
Various components of the pertinent error budget are discussed
in detail -- starting from the context of the LEP experiments, 
through their current updates, up to prospects 
of their improvements for the sake of the FCC-ee. 
It is argued that with an appropriate upgrade of the Monte
Carlo event generator \bhlumi\ and/or other similar MC programs
calculating QED effects in the low angle Bhabha process,
the total theoretical error 
of $0.01\%$ for the FCC-ee luminometry can be reached.
A new study of the $Z$ and $s$-channel $\gamma$ exchanges within
the angular range of the FCC-ee luminometer
using the \bhwide\ Monte Carlo was instrumental in obtaining the above result.
Possible ways of \bhlumi\ upgrade are also discussed.
\end{abstract}

\vspace{40mm}
\footnoterule
\noindent
{\footnotesize
$^{\star}$This work is partly supported by
%%%%% NCN-Jadach
 the Polish National Science Center grant 2016/23/B/ST2/03927
% the Citadel Foundation 
 and the CERN FCC Design Study Programme.
}

\end{titlepage}

\section{Introduction}
%%%%%%%%%%%%%%%%%%%%%%%%%%
The current status of the theoretical precision for the Bhabha
luminometry is critically reviewed and pathways are outlined
to the requirement targeted by the FCC-ee precision studies.
Various components of the pertinent error budget are discussed
in detail -- starting from the context of the LEP experiments, 
through their current updates, up to prospects 
of their improvements for the sake of the FCC-ee. 
It is argued that, with an appropriate upgrade of the Monte
Carlo event generator \bhlumi\ and/or other similar MC programs
calculating QED effects in 
the low angle Bhabha (LABH) process $e^-e^+ \to e^-e^+$,
the total theoretical error 
of $0.01\%$ for the luminometry at the high luminosity FCC-ee
machine~\cite{Gomez-Ceballos:2013zzn} can be reached. 
Possible ways of this upgrade are also discussed.

In Section~\ref{sec:lumi2} we recap the main aspects of
the theoretical precision in the LEP luminosity measurement
and present important components of the corresponding error budget.
In Section~\ref{sec:lumi3} we present current improvements
on some of the above components.
In Section~\ref{sec:lumi4} we discuss in detail prospects on
reaching the $0.01\%$ theory precision for the FCC-ee
luminometry and outline ways of upgrading the main Monte 
Carlo program for this purpose, \bhlumi, in this respect.
In Section~\ref{sec:tech-prec} the important issue of technical precision 
is addressed.
Finally, in Section~\ref{sec:lumi5} we briefly summarize our work.

%%%%%%%%%%%%%%%%%%%%%%%%%%%%%%%%%%%%%%%%%%%%%%%%%%%%%%%%%%%%%%%%%
\section{Theoretical uncertainty in LEP luminometry, A.D. 1999}
\label{sec:lumi2}
Let us recapitulate the essential aspects of the theory (mainly QED)
uncertainty in the LEP luminometry, as seen A.D. 1999.
Luminosity measurements of all four LEP collaborations at CERN 
and also of SLD at SLAC
relied on theoretical predictions for the low-angle
Bhabha process obtained using the \bhlumi\ Monte Carlo multiphoton event generator
featuring a sophisticated QED matrix element with soft photon resummation.
Its version 2.01 was published in 1992 (see ref.~\cite{Jadach:1991by})
and the upgraded version 4.04 was published in ref~\cite{Jadach:1996is}.

%%%%%%%%%%%%%%%%%%%%%%%%%%%%%%%%%%%%%%%%%%%%%%%%%%%%%%%%%%%%%%%%%%%%%%%%%%%%%%%
%%%%%%%%%%%%%%%%%%%%%%%%%%%%%%%%%%%%%%%%%%%%%%%%%%%%%%%%%%%%%%%%%%%%%%%%%%%%%%%
\begin{table}[!ht]
\centering
\begin{tabular}{|l|l|l|l|l|l|}
\hline %%%%%%%%%%%%%%%%%%%%%%%%%%%%%%%%%%%%%%%%%%%%%%%%%%%%%%%%%%%%%%
    & \multicolumn{2}{|c|}{LEP1} 
              & \multicolumn{2}{|c|}{LEP2}
\\ \hline %%%%%%%%%%%%%%%%%%%%%%%%%%%%%%%%%%%%%%%%%%%%%%%%%%%%%%%%%%%
Type of correction/error
    & 1996
         & 1999
              & 1996
                   & 1999
\\  \hline %%%%%%%%%%%%%%%%%%%%%%%%%%%%%%%%%%%%%%%%%%%%%%%%%%%%%%%%%%
%Technical Precision& -- & --& --& (0.03)\% 
%\\ %%%%%%%%%%%%%%%%%%%%%%%%%%%%%%%%%%%%%%%%%%%%%%%%%%%%%%%%%%%%%%%%%%
(a) Missing photonic ${\cal O}(\alpha^2 )$~\cite{Jadach:1995hy,Jadach:1999pf} 
    & 0.10\%      
        & 0.027\%    
            & 0.20\%  
                & 0.04\%
\\ %%%%%%%%%%%%%%%%%%%%%%%%%%%%%%%%%%%%%%%%%%%%%%%%%%%%%%%%%%%%%%%%%%
(b) Missing photonic ${\cal O}(\alpha^3 L_e^3)$~\cite{Jadach:1996ir} 
    & 0.015\%     
        & 0.015\%    
            & 0.03\%  
                & 0.03\% 
\\ %%%%%%%%%%%%%%%%%%%%%%%%%%%%%%%%%%%%%%%%%%%%%%%%%%%%%%%%%%%%%%%%%%
(c) Vacuum polarization~\cite{Burkhardt:1995tt,Eidelman:1995ny} 
    & 0.04\%      
        & 0.04\%    
           & 0.10\%  
                & 0.10\% 
\\ %%%%%%%%%%%%%%%%%%%%%%%%%%%%%%%%%%%%%%%%%%%%%%%%%%%%%%%%%%%%%%%%%%
(d) Light pairs~\cite{Jadach:1992nk,Jadach:1996ca} 
    & 0.03\%      
        & 0.03\%    
            & 0.05\%  
                & 0.05\% 
\\ %%%%%%%%%%%%%%%%%%%%%%%%%%%%%%%%%%%%%%%%%%%%%%%%%%%%%%%%%%%%%%%%%%
(e) $Z$ and $s$-channel $\gamma$~\cite{Jadach:1995hv,Arbuzov:1996eq}
    & 0.015\%      
        & 0.015\%   
            &  0.0\%  
                & 0.0\% 
\\ \hline %%%%%%%%%%%%%%%%%%%%%%%%%%%%%%%%%%%%%%%%%%%%%%%%%%%%%%%%%%%
Total  
    & 0.11\%~\cite{Arbuzov:1996eq}
        & 0.061\%~\cite{Ward:1998ht}
            & 0.25\%~\cite{Arbuzov:1996eq}
                & 0.12\%~\cite{Ward:1998ht}
\\ \hline %%%%%%%%%%%%%%%%%%%%%%%%%%%%%%%%%%%%%%%%%%%%%%%%%%%%%%%%%%%
\end{tabular}
%30Aug \setcounter{table}{0}
\caption{\sf
Summary of the total (physical+technical) theoretical uncertainty
for a typical calorimetric detector.
For LEP1, the above estimate is valid for a generic angular range
within   $1^{\circ}$--$3^{\circ}$ ($18$--$52$ mrads), and
for  LEP2 energies up to $176$~GeV and an
angular range within $3^{\circ}$--$6^{\circ}$.
Total uncertainty is taken in quadrature.
Technical precision included in (a).
}
\label{tab:error99}
\end{table}
%%%%%%%%%%%%%%%%%%%%%%%%%%%%%%%%%%%%%%%%%%%%%%%%%%%%%%%%%%%%%%%%%%%%%%%%%%%%%%%

The theoretical uncertainty of the \bhlumi\ Bhabha prediction,
initially rated at 0.25\% \cite{Jadach:1991cg}, 
was re-evaluated in 1996 after extensive tests and debugging 
to be 0.16\%~\cite{Jadach:1995pd}.
From that time, the code of \bhlumi\ version 4.04
used by all LEP collaborations in their data analysis remains frozen.
The following re-evaluation of its precision came from investigations
using external calculations outside the \bhlumi\ code.
For instance, the 0.11\% estimate of ref.~\cite{Arbuzov:1996eq}
was based on better estimations of the QED corrections missing in \bhlumi\ 
and on improved knowledge of the vacuum polarization contribution.
The detailed composition of the final estimate of the theoretical uncertainty
$\delta \sigma / \sigma \simeq 0.061\%$
of the \bhlumi~4.04 prediction, based on published works, 
is shown in Table~\ref{tab:error99}, following ref.~\cite{Ward:1998ht}.
This value was used in the final LEP1 data analysis
in ref.~\cite{ALEPH:2005ab}.
On the other hand, at LEP2 the experimental error was substantially
larger than the QED uncertainty of the Bhabha process listed 
in Table~\ref{tab:error99}, where we define $L_e=\ln(|t|/m_e^2)$.

All four LEP collaborations were quoting the experimental luminosity errors 
for LEP1 data below 0.05\%, that is below the theoretical error.
The best experimental luminosity error 0.034\%
was quoted by the OPAL collaboration%
\footnote{The OPAL collaboration has found all their experimental
   distributions for low-angle Bhabha data to be in a striking agreement 
   with the \bhlumi\ Monte Carlo simulation~\cite{Abbiendi:1999zx}.}
-- they also quoted a slightly smaller theory error, 0.054\%, 
thanks to use of improved light-fermion-pair calculations of
refs.~\cite{Montagna:1998vb,Montagna:1999eu};
see also the review article~\cite{CarloniCalame:2015zev}
and workshop presentations~\cite{jadach:2006fcal,CarloniPisa:2015}.

%%%%%%%%%%%%%%%%%%%%%%%%%%%%%%%%%%%%%%%%%%%%%%%%%%%%%%%%%%%%%%%%%%%%%%%%%%
\section{Present status (2018)}
\label{sec:lumi3}
%%%%%%%%%%%%%%%%%%%%%%%%%%%%%%%%%%%%%%%%%%%%%%%%%%%%%%%%%%%%%%%%%%%%%%%%%%
From the end of LEP until the present time there has been limited progress
on practical calculations for low-angle Bhabha
scattering at energies around and above the $Z$ resonance%
\footnote{
 This is in spite of a considerable effort on the
 ${\cal O}(\alpha^2)$ so-called "fixed-order" 
 (without resummation) QED calculations for the Bhabha process; 
 see below for more discussion.}.
A new Monte Carlo generator \babayaga\ based on the parton shower
algorithm was developed%
\cite{CarloniCalame:2000pz,CarloniCalame:2001ny,Balossini:2006wc,CarloniCalame:2015zev}. 
It was intended mainly for low energy electron--positron colliders 
with $\sqrt{s} \leqslant 10$~GeV, claiming precision at $0.1\%$,
but was not validated for energies near the $Z$ peak.

There was, however, a steady improvement in the precision 
of the vacuum polarization in the $t$-channel photon propagator;
see the recent review in the FCC-ee workshop~\cite{JegerlehnerCERN:2016}.
Using the uncertainty $\delta\Delta^{(5)}_{had.}=0.63\cdot 10^{-4}$
at $\sqrt{-t}=2$GeV quoted in Ref.~\cite{Jegerlehner:2017zsb} one obtains 
$\delta \sigma/\sigma =1.3 \cdot 10^{-4}$.
It is shown in the second column
in Table~\ref{tab:lep-update}, marked "Update 2018".
The improvement of the light-pair corrections of
refs.~\cite{Montagna:1998vb,Montagna:1999eu}
is also taken into account there.

%%%%%%%%%%%%%%%%%%%%%%%%%%%%%%%%%%%%%%%%%%%%%%%%%%%%%%%%%%%%%%%%%%%
\begin{table}[ht!]
\centering
\begin{tabular}{|l|l|l|l|}
\hline
Type of correction~/~Error
    &  1999
        & Update 2018
\\ \hline %%%%%%%%%%%%%%%%%%%%%%%%%%%%%%%%%%%%%%%%%%%%%%%%%%%%%%%%%%%
(a) Photonic ${\cal O}(L_e\alpha^2 )$
    & 0.027\% ~\cite{Jadach:1999pf}
        & 0.027\%
\\ %%%%%%%%%%%%%%%%%%%%%%%%%%%%%%%%%%%%%%%%%%%%%%%%%%%%%%%%%%%%%%%%%%
(b) Photonic ${\cal O}(L_e^3\alpha^3)$
    & 0.015\%~\cite{Jadach:1996ir}
        & 0.015\%
\\ %%%%%%%%%%%%%%%%%%%%%%%%%%%%%%%%%%%%%%%%%%%%%%%%%%%%%%%%%%%%%%%%%%
(c) Vacuum polariz.
    &0.040\%~\cite{Burkhardt:1995tt,Eidelman:1995ny} 
        & 0.013\%~\cite{JegerlehnerCERN:2016}
\\ %%%%%%%%%%%%%%%%%%%%%%%%%%%%%%%%%%%%%%%%%%%%%%%%%%%%%%%%%%%%%%%%%%
(d) Light pairs
    & 0.030\%~\cite{Jadach:1996ca}
        & 0.010\%~\cite{Montagna:1998vb,Montagna:1999eu}
\\ %%%%%%%%%%%%%%%%%%%%%%%%%%%%%%%%%%%%%%%%%%%%%%%%%%%%%%%%%%%%%%%%%%
(e) $Z$ and $s$-channel $\gamma$ exchange
    &0.015\%~\cite{Jadach:1995hv,Arbuzov:1996eq}
        & 0.015\%
\\ %%%%%%%%%%%%%%%%%%%%%%%%%%%%%%%%%%%%%%%%%%%%%%%%%%%%%%%%%%%%%%%%%%
(f) Up-down interference
    &0.0014\%~\cite{Jadach:1990zf}
        & 0.0014\%
\\ %%%%%%%%%%%%%%%%%%%%%%%%%%%%%%%%%%%%%%%%%%%%%%%%%%%%%%%%%%%%%%%%%%
(f) Technical Precision& -- & (0.027)\%
\\ \hline  %%%%%%%%%%%%%%%%%%%%%%%%%%%%%%%%%%%%%%%%%%%%%%%%%%%%%%%%%%
Total
    & 0.061\%~\cite{Ward:1998ht}
        & 0.038\%
\\ \hline  %%%%%%%%%%%%%%%%%%%%%%%%%%%%%%%%%%%%%%%%%%%%%%%%%%%%%%%%%%
\end{tabular}
\caption{\sf
Summary of the total (physical+technical) theoretical uncertainty for a typical
calorimetric LEP luminosity detector within the generic angular range
of $18$--$52$\,mrad.
Total error is summed in quadrature.
}
\label{tab:lep-update}
\end{table}
%%%%%%%%%%%%%%%%%%%%%%%%%%%%%%%%%%%%%%%%%%%%%%%%%%%%%%%%%%%%%%%%%%%%
%%%%%%%%%%%%%%%%%%%%%%%%%%%%%%%%%%%%%%%%%%%%%%%%%%%%%%%%%%%%%%%%%%%%

The important point is that the technical precision,
which is marked in parentheses as $0.027\%$, is not included in the sum,
because according to ref.~\cite{Arbuzov:1996eq} it is
included in the uncertainty of the photonic corrections.
Future reduction of the photonic correction error will require a
clear separation of the technical precision from other uncertainties
and it may turn out to be a dominant one.

%%%%%%%%%%%%%%%%%%%%%%%%%%%%%%%%%%%%%%%%%%%%%%%%%%%%%%%%%%%%%%%%%%%
\section{Path to 0.01\% precision for FCC-ee}
\label{sec:lumi4}

In the following we shall describe what steps are needed on
the path to the $\leq 0.01\%$ precision required
for the low-angle Bhabha (LABH) luminometry at the FCC-ee experiments.
The last column in Table~\ref{tab:lep2fcc} summarizes 
this goal component-by-component
in the precision forecast for the FCC-ee luminometry.
We will also specify all improvements
in the next version \bhlumi\ which could bring us to 
the FCC-ee precision level.

%%%%%%%%%%%%%%%%%%%%%%%%%%%%%%%%%%%%%%%%%%%%%%%%%%%%%%%%%%%%%%%%%%%%
\begin{table}[ht!]
\centering
\begin{tabular}{|l|l|l|l|}
\hline
Type of correction~/~Error
        & Update 2018
                &  FCC-ee forecast
\\ \hline %%%%%%%%%%%%%%%%%%%%%%%%%%%%%%%%%%%%%%%%%%%%%%%%%%%%%%%%%%
(a) Photonic $[{\cal O}(L_e\alpha^2 )]\; {\cal O}(L_e^2\alpha^3)$
        & 0.027\%
                &  $ 0.1 \times 10^{-4} $
\\ %%%%%%%%%%%%%%%%%%%%%%%%%%%%%%%%%%%%%%%%%%%%%%%%%%%%%%%%%%%%%%%%%
(b) Photonic $[{\cal O}(L_e^3\alpha^3)]\; {\cal O}(L_e^4\alpha^4)$
        & 0.015\%
                & $ 0.6 \times 10^{-5} $
\\%%%%%%%%%%%%%%%%%%%%%%%%%%%%%%%%%%%%%%%%%%%%%%%%%%%%%%%%%%%%%%%%%
(c) Vacuum polariz.
        & 0.014\%~\cite{JegerlehnerCERN:2016}
                & $ 0.6 \times 10^{-4} $
\\%%%%%%%%%%%%%%%%%%%%%%%%%%%%%%%%%%%%%%%%%%%%%%%%%%%%%%%%%%%%%%%%%%
(d) Light pairs
        & 0.010\%~\cite{Montagna:1998vb,Montagna:1999eu}
                & $ 0.5 \times 10^{-4} $
\\%%%%%%%%%%%%%%%%%%%%%%%%%%%%%%%%%%%%%%%%%%%%%%%%%%%%%%%%%%%%%%%%%%
(e) $Z$ and $s$-channel $\gamma$ exchange
        & 0.090\%~\cite{Jadach:1995hv}
                & $ 0.1 \times 10^{-4} $
\\ %%%%%%%%%%%%%%%%%%%%%%%%%%%%%%%%%%%%%%%%%%%%%%%%%%%%%%%%%%%%%%%%%%
(f) Up-down interference
    &0.009\%~\cite{Jadach:1990zf}
        & $ 0.1 \times 10^{-4} $
\\%%%%%%%%%%%%%%%%%%%%%%%%%%%%%%%%%%%%%%%%%%%%%%%%%%%%%%%%%%%%%%%%%
(f) Technical Precision & (0.027)\% 
                & $ 0.1 \times 10^{-4} $
\\ \hline %%%%%%%%%%%%%%%%%%%%%%%%%%%%%%%%%%%%%%%%%%%%%%%%%%%%%%%%%
Total
        & 0.097\%
                & $ 1.0 \times 10^{-4} $
\\ \hline %%%%%%%%%%%%%%%%%%%%%%%%%%%%%%%%%%%%%%%%%%%%%%%%%%%%%%%%%%
\end{tabular}
\caption{\sf
Anticipated total (physical+technical) theoretical uncertainty 
for a FCC-ee luminosity calorimetric detector with
the angular range being $64$--$86\,$mrad (narrow), near the $Z$ peak.
Description of photonic corrections in square brackets is related to 
the 2nd column.
The total error is summed in quadrature.
}
\label{tab:lep2fcc}
\end{table}
%%%%%%%%%%%%%%%%%%%%%%%%%%%%%%%%%%%%%%%%%%%%%%%%%%%%%%%%%%%%%%%%%%%
%%%%%%%%%%%%%%%%%%%%%%%%%%%%%%%%%%%%%%%%%%%%%%%%%%%%%%%%%%%%%%%%%%%

Before coming to the details of the envisaged improvements in QED
calculations for the LABH process,
let us recapitulate briefly basic features of the LABH luminometry
which have to be kept in mind in QED perturbative calculations for FCC-ee.
First of all, the largest photonic QED effects due to multiple real
and virtual photon emission are strongly cut-off dependent.
Event acceptance of the LABH luminometer is quite complicated,
and cannot be dealt with analytically, 
hence a Monte Carlo implementation of QED perturbative results is mandatory.
The LABH detector at FCC-ee will be similar to that of LEP, with calorimetric
detection of electrons and photon (not distinguishing them)
within the angular range $(\theta_{\min},\theta_{\max})$ on opposite
sides of the collision point~\cite{MogensDam:2018}.
The detection rings are divided into small cells and the angular range
on both sides is slightly different in order to minimize QED effects.
The angular range at FCC-ee is planned 
to be $64$--$86\,$mrads (narrow)~\cite{MogensDam:2018}
while at LEP it was typically $28$--$50\,$mrads
(narrow range, ALEPH/OPAL silicon detector);
see Fig.~2 in ref.~\cite{Jadach:1995pd}
(also Fig.~16 in ref.~\cite{Jadach:1996gu}) 
for an idealized detection algorithm of the generic LEP silicon detector.
The average $t$-channel transfer near the $Z$ resonance will be 
$|\bar{t}|^{1/2} = \langle |t| \rangle^{1/2} \simeq 3.25$\,GeV at FCC-ee
instead of 1.75\,GeV at LEP.%
\footnote{ At $350\,$GeV, the FCC-ee luminometer will have $|\bar{t}| = 12.5\,$GeV.}
The important scale factor controlling photonic QED effects,
$\gamma=\frac{\alpha}{\pi}\ln\frac{|\bar{t}|}{m_e^2}=0.042$
for FCC-ee, that is only slightly greater than $0.039$ for LEP.
On the other hand, the factor $x=|t|/s$ suppressing
$s$-channel contributions will be $1.27\times 10^{-3}$,
significantly larger than $0.37 \times 10^{-3}$ for LEP.

Finally, let us remark that the process $e^+e^-\to 2\gamma$
is also considered for FCC-ee luminometry, 
see refs.~\cite{CarloniCalame:2015zev,Balossini:2008xr} 
for more discussion on the QED radiative corrections to this process.

\subsection{Photonic higher-order and subleading corrections}
%%%%%%%%%%%%%%%%%%%%%%%%%%%%%%%%%%%%%%%%%%%%%%%%%%%%%%%%%%%%%
Photonic corrections (items (a) and (b) in Table~\ref{tab:lep-update})
are large but they are mainly due to collinear and soft
singularities which are known in QED at any perturbative order,
hence can be resummed.
The cross section of the the LABH luminometer is highly sensitive
to emission of real soft and collinear photons.
Even relatively soft collinear photon emission in the initial state (ISR)
may pull final electrons outside the acceptance angular range,
while final-state photons can easily change the shape of the final
state ``calorimetric cluster".
This is why resummation of the multiple photon effects has
to be implemented in an exclusive way, using the method of
exclusive exponentiation (EEX), as in \bhlumi~\cite{Jadach:1996is},
or using the parton shower (PS) method 
as in \babayaga~\cite{CarloniCalame:2000pz}.
It was shown~\cite{slac-talk} that, for instance, 
the so-called ``fixed-order"
\order{\alpha^2} calculations without resummation%
\footnote{For instance, see the calculations of 
  refs.~\cite{Penin:2005eh,Czakon:2005gi}.}
are completely inadequate for the LABH luminometry, leaving out uncontrolled
QED effects of the order $\sim 0.5\%$ in the angular
distribution, and even a few percent in some other important distributions.

Assuming that the technical precision is dealt with separately
(see the discussion in the following Section~\ref{sec:tech-prec}),
item (a) in Table~\ref{tab:lep-update},
missing in \bhlumi\ v. 4.04, scales like $L_e=\ln(|t|/m_e^2)$,
where $t$ is the relevant squared momentum transfer.
However, this item will disappear from the error budget completely
once the EEX matrix element of \bhlumi\ is upgraded to include
\order{L_e\alpha^2} contributions, which are already known and published.
In fact, these \order{L_e \alpha^2} corrections 
consist of 2-real photon contributions,
1-loop corrections to 1-real emission and 2-loop corrections.
Efficient numerical and analytic methods of calculating 
the exact \order{\alpha^2} matrix element (spin amplitudes)
for 2 real photons, keeping fermion masses, have been known for decades; 
see refs.~\cite{Kleiss:1985yh,Berends:1984qf}.
In ref.~\cite{Jadach:1992tf} exact 2-photon amplitudes were compared
with the matrix element of \bhlumi.

Truly pioneering work on \order{L_e \alpha^2, L_e^0 \alpha^2} 
virtual corrections to 1-photon distributions was done in ref.~\cite{Jadach:1995hy}.
These were calculated neglecting interference terms between $e^+$ and $e^-$
lines, which near the $Z$ peak are of the order of
$\big(\frac{\alpha}{\pi}\big)^2  \frac{|t|}{s} L_e \sim 10^{-7}$ 
times some logarithm of the cut-off.
Let us note in passing that we know from the $s$-channel analog in \cite{Jadach:2006fx}
that the pure \order{L_e \alpha^2} 
correction of this class (neglecting the \order{L_e^0 \alpha^2} term)
is amazingly compact --
it consists of merely a 3-line formula at the amplitude level.
Let us add for completeness that the above correction
was also calculated numerically in ref.~\cite{Actis:2009uq}.

Finally, in ref.\cite{Ward:1998ht}, the
two-loop \order{L_e \alpha^2} $t$-channel photon form-factor relevant
for the LABH process (keeping in mind $|t|/s$ suppression) continued
analytically from the known $s$-channel result of ref.~\cite{Berends:1987ab}
was added, thus accounting for the complete \order{L_e \alpha^2}
photonic correction, known but not included in the MC \bhlumi\ v4.04.
Once the above well-known photonic \order{L_e \alpha^2} 
part is added in the future upgrade of the EEX matrix element in \bhlumi, 
the corresponding item will disappear from the list of 
the projected FCC-ee luminometry uncertainties
in Table.~\ref{tab:lep2fcc}.

In view of the above discussion, it is clear that the major effort
of calculating the complete \order{\alpha^2} QED correction to
low and wide-angle Bhabha processes in 
refs.~\cite{Czakon:2005gi, Penin:2005eh, Bern:2000ie,Bonciani:2005im},
see also \cite{Actis:2007gi,Actis:2008br,Bonciani:2007eh,Kuhn:2008zs},
is of rather limited practical importance for the LABH luminometry at FCC-ee%
\footnote{They are more relevant for the wide-angle Bhabha, 
  provided they are included in the MC with soft-photon resummation.
  However, this is rather problematic, because in all these works
  soft-real-photon contributions are added to loop corrections
  {\em a la} Bloch--Nordsieck, instead of subtracting the well-known
  virtual form-factor from virtual loop results already
  at the amplitude level, before squaring them.
  }.
All these works essentially add previously unknown 
\order{L_e^0 \alpha^2} corrections,
which are of order $\sim 10^{-5}$.
Their size should be checked%
\footnote{This kind of correction is often enhanced by $\pi^2$ factors.}
using auxiliary programs outside the \bhlumi\, Monte Carlo,
in order to be listed among QED uncertainties in the uncertainty budget
as in our Table.~\ref{tab:lep2fcc}.
In any case, we expect corrections of this class to stay 
well below $10^{-4}$,
and most likely there will be no need to add the complete
\order{L_e^0 \alpha^2} corrections to the matrix element of any MC
for the LABH process.

Another important photonic correction listed
as item (b) in Table~\ref{tab:error99}
as an uncertainty of \bhlumi\ is the
\order{\alpha^3 L_e^3} correction (third order LO).
It is already known from Ref.~\cite{Jadach:1996bx,Jadach:1996ir} 
and is currently omitted from v. 4.04 of \bhlumi,
although already included in the LUMLOG part of \bhlumi.
Given its already known size, we would need to implement 
this  third order leading-order result into 
the EEX matrix element of \bhlumi,
and it will disappear from the uncertainty list.
Once it is done, the uncertainty due to 
\order{\alpha^4 L_e^4} and \order{\alpha^3 L_e^2}
should be estimated and included in the list of photonic
uncertainties of \bhlumi\, with the upgraded EEX matrix element.
We can use the scaling rules indicated in the previous discussion
to estimate an error due to missing \order{\alpha^4 L_e^4}
as $0.015\% \times \gamma= 0.6\times 10^{-5}$ near the $Z$ peak.
The scale of the missing \order{\alpha^3 L_e^2} 
is also of a similar order, $\gamma^2 \alpha/\pi \simeq 10^{-5}$,
and its actual estimate is currently highly uncertain.

The so called up-down interference between photon emission from $e^+$ 
and $e^-$ lines was  calculated in ref.~\cite{Jadach:1990zf}
at \order{\alpha^1} to be roughly $\delta\sigma/\sigma \simeq 0.07\; |t|/s$.
At LEP1 its contribution is negligible, see Table~\ref{tab:lep-update},
but at the FCC-ee luminometer it will be the factor of 10 larger and has to be included
in the matrix element of the upgraded \bhlumi.
Once it is done, its uncertainty should be again negligible,
as indicated in Table~\ref{tab:lep2fcc}, 
where we used $ 2\gamma \times 0.07\; |t|/s $ as a crude estimator 
of its future uncertainty.

\subsection{EEX versus CEEX matrix element}
%%%%%%%%%%%%%%%%%%%%%%%%%%%%%%%%%%%%%%%%%%%%%%%%%%%%
\bhlumi\ multi-photon distributions obey a clear separation
into exact Lorentz invariant phase space and squared matrix element.
The matrix element is an independent part of the program and is currently
built according to exclusive exponentiation (EEX) based on the
Yennie--Frautshi--Suura~\cite{Yennie:1961ad} (YFS)
soft photon factorization and resummation
performed on the spin-summed squared amplitude.
It includes complete \order{\alpha^1}
and \order{L_e^2 \alpha^2} corrections,
neglecting interference terms between electron and positron lines,
suppressed by a $|t|/s$ factor.

Let us underline that the above EEX-style
matrix element in \bhlumi\ has not been changed in the upgrades 
since version 2.01~\cite{Jadach:1991by}.
As already said, we may continue this practice 
and introduce the results from 
Refs.~\cite{Jadach:1995hy,Jadach:1996bx,Jadach:1996ir,Ward:1998ht}
into the EEX matrix element, that is \order{\alpha^2 L_e}
and \order{\alpha^3 L_e^3}, neglecting again some $\sim|t|/s$ terms.

On the other hand, using the same underlying 
multi-photon phase space
MC generator of \bhlumi\ and exploiting 
the results from Refs.~\cite{Jadach:1995hy,Jadach:1996bx,Jadach:1996ir,Ward:1998ht}, 
one could implement a more sophisticated matrix element of
the CEEX~\cite{Jadach:2000ir} type,
where CEEX stands for coherent exclusive exponentiation.
In the CEEX resummation methodology, soft photon factors
are factorized at the amplitude level and the matching
with fixed order results is also done
at the amplitude level (before squaring and spin-summing).
The big advantage of CEEX over EEX is that the separation of the infrared (IR)
parts and matching with the fixed-order result are much simpler and 
more transparent when done at the amplitude level -- all IR cancellations for
complicated interferences are managed automatically and numerically.
The inclusion of the $s$-channel $Z$ and photon exchange and $t$-channel $Z$
exchange including \order{\alpha} corrections, soft photon
interference between electron and positron lines, and all that
would be much easier to take into account for CEEX than in the case of EEX.
However, the inclusion of \order{\alpha^3 L_e^3} in CEEX will have to
be worked out and implemented.

Summarizing, the CEEX version would allow a more systematic 
further development 
of the program as we move forward with the FCC-ee project. 
From this perspective, the CEEX version is preferable,
although the improvement of the EEX matrix element 
should be also pursued. 
See some additional discussion in Sect.~\ref{sec:tech-prec}.

\subsection{Error on hadronic vacuum-polarization contribution}
%%%%%%%%%%%%%%%%%%%%%%%%%%%%%%%%%%%%%%%%%%%%%%%%%%%%%%%%%%%%%%%%%%%
The uncertainty of the low-angle Bhabha cross section
due to imprecise knowledge of the QED running coupling constant
of the $t$-channel photon exchange is simply
$\frac{\delta_{VP}\sigma}{\sigma}
= 2\frac{\delta\alpha_{eff}(\bar{t})}{\alpha_{eff}(\bar{t})}$,
where $\bar{t}$ is the average transfer of the $t$-channel photon.
For the FCC-ee luminometer, it will be $|\bar{t}|^{1/2} \simeq 3.5\,$GeV
near the Z peak and $|\bar{t}|^{1/2} \simeq  13\,$GeV at $350\,$GeV.

The uncertainty of $\alpha_{\rm eff}(t)$ is mainly due to the use of
the experimental cross section $\sigma_{\rm had}$
for $e^-e^+\to hadrons$ below $10\,$GeV
as an input to the (subtracted) dispersion relations.
A comprehensive review of the corresponding methodology
and the latest update of the results can be found
in refs.~\cite{Jegerlehner:2006ju,Jegerlehner:2017gek}, 
see also the FCC-ee workshop presentation~\cite{JegerlehnerCERN:2016}.

In the above works, the hadronic contribution to $\alpha_{\rm eff}$
from the dispersion relation is encapsulated in 
$\Delta \alpha^{(5)}(-s_0)$, where $2\,$GeV$\leq s_0^{1/2} \leq 10\,$GeV
in order to minimize the dependence on $\sigma_{had}(s)$,
such that the main contribution comes from $s^{1/2}\leq 2\,$GeV.
Moreover, prospects of improving experimental
data on $\sigma_{\rm had}(s)$ in this energy range are very good 
also, because the main contribution to the error in the measurement 
of the muon $g-2$ comes from the same cross section range~\cite{Jegerlehner:2006ju}.

The above works are focusing on the parameter range 
$2\,$GeV$\leq s_0^{1/2} \leq 10\,$GeV,
which is accidentally of paramount interest for the FCC-ee luminometry,
are part of a wider strategy in
refs.~\cite{Jegerlehner:2006ju,Jegerlehner:2017gek}
of obtaining $\alpha_{\rm eff}(M_Z^2)$ in two steps,
where $\Delta \alpha^{(5)}(-s_0)$ is obtained from dispersion relations
and the difference $\Delta \alpha^{(5)}(M_Z^2) - \Delta \alpha^{(5)}(-s_0)$
is calculated using the perturbative QCD technique of 
the Adler function~\cite{Eidelman:1998vc}.
The error of the above difference due to limited knowledge of $\alpha_s$,
the $c$ and $b$ quark masses and higher-order perturbative QCD effects
is small enough, such that
the overall uncertainty of $\alpha_{\rm eff}(M_Z^2)$ 
is smaller than that from the direct use of the dispersion relation.

Taking $s_0^{1/2} = 2.0\,$GeV and the value 
$\Delta \alpha^{(5)} (-s_0) = (64.09 \pm 0.63) \times 10^{-4},\; $
of ref.~\cite{Jegerlehner:2017zsb} as a benchmark,
in Table~\ref{tab:lep-update} we quote 
$(\delta_{\rm VP} \sigma)/\sigma = 1.3 \times 10^{-4}$.
Thanks to anticipated improvements of data for 
$\sigma_{\rm had}(s)$, $s^{1/2}\leq 2.5\,$GeV,
one may expect the factor of $2$ improvement by the time of the FCC-ee experiments,
that is $\delta_{\rm VP} \sigma/\sigma = 0.65 \times 10^{-4}$ 
near the $Z$ peak, see Table~\ref{tab:lep2fcc}.

At the high-energy end of FCC-ee, $350\,$GeV,
due to the increase of the average transfer $|\bar{t}|=12.5\,$GeV,
one obtains presently from the dispersion relation
$\delta \alpha_{\rm eff}/\alpha_{\rm eff}=1.190\times 10^{-4}$
and $\delta_{\rm VP} \sigma/\sigma \simeq 2.4 \times 10^{-4}$,
and again with the possible improvement of the factor of $2$, 
so that the FCC-ee expectation%
\footnote{We thank F.\ Jegerlehner for elucidating
         private communications on the above predictions.}
is $(\delta_{VP} \sigma)/\sigma \simeq 1.2 \times 10^{-4}$.

There are also alternative proposals for 
the measurement of $\alpha_{\rm eff}(t)$
not relying (or relying less) on dispersion relations;
see refs.~\cite{Janot:2015gjr,Abbiendi:2016xup}.
Ref.~\cite{Janot:2015gjr} proposed a method for the direct
measurement of $\alpha_{\rm eff}(M_Z^2)$ using charge asymmetry
in $e^-e^+\to \mu^-\mu^+$ near the $Z$ resonance.
One may ask whether its precise value can also be used
to predict very precisely $\alpha_{\rm eff}(t)$
in the FCC-ee luminometer range $2\,$GeV$\leq |t|^{1/2} \leq 10\,$GeV?
It turns out that the uncertainty due to the use 
of pQCD~\cite{JegerlehnerCERN:2016}
in the transition from the $M_Z$ scale 
down to below $10\,$GeV is about the same as in the traditional methods.
However, a direct measurement of $\alpha_{\rm eff}(M_Z^2)$
may serve as an important crosscheck.
The other proposal, in ref.~\cite{Abbiendi:2016xup},
of the direct measurement of $\alpha_{\rm eff}(t),\; t\sim -1\,$GeV$^2$,
from the elastic scattering of energetic muons on atomic
electrons sounds interesting, but requires more studies.

\subsection{The uncertainty due to light fermion pairs}
%%%%%%%%%%%%%%%%%%%%%%%%%%%%%%%%%%%%%%%%%%%%%%%%%%%%%%%%%%%%%

Three groups of calculations are available for 
the light-fermion-pair effect in the low angle Bhabha process:
\cite{Jadach:1992nk,Jadach:1996ca},
\cite{Montagna:1998vb,Montagna:1999eu} and 
\cite{Arbuzov:1996zp, Arbuzov:1995qd, Arbuzov:1995cn, 
Arbuzov:1996su,Merenkov:1997zm}.

The biggest correction, due to additional electron pair production,
was calculated in Ref.~\cite{Montagna:1998vb},
where process $e^+e^-\to e^+e^-e^+e^-$ calculated
with the help of the ALPHA algorithm \cite{Caravaglios:1995cd}
was combined with virtual/soft corrections of
Refs.~\cite{Barbieri:1972as,Barbieri:1972hn,Burgers:1985qg},
resulting in the theoretical error on pair correction to be 0.01\%.
\footnote{
The emission of a $\mu$-pair is also discussed in Ref.~\cite{Montagna:1998vb}
}
This value is quoted in Table~\ref{tab:lep-update}
as the present state of the art for the uncertainty of corrections
due to light fermion pair production.

In Refs.~\cite{Arbuzov:1995qd, Arbuzov:1995cn} $e^+e^-$ pair corrections 
were calculated in a semi-analytic way at NLO accuracy, 
omitting non-logarithmic corrections
and taking virtual corrections from \cite{Barbieri:1972as,Barbieri:1972hn}. 
The third order LO correction due to simultaneous emission of
the additional $e^+e^-$ pair (Non-Singlet and Singlet) and additional photon
were also evaluated.
The overall precision of the Bhabha scattering formula of 
Refs.~\cite{Arbuzov:1995qd, Arbuzov:1995cn} was estimated there to be $0.006\%$,
mainly due to omission of the heavier lepton pairs 
($\mu^+\mu^-$, $\tau^+\tau^-$) and quark pairs ($0.005\%$). 
One can assume conservatively the same $0.006\%$ as the total error on additional 
pair correction. 

In the Ref.~\cite{Jadach:1992nk} the complete LO semi-analytic 
calculations based on the electron structure functions 
were presented up to the third order for the Non-Singlet%
\footnote{This is contrary to the incorrect statement 
  in Ref.~\cite{Arbuzov:1995qd}. 
  Third order NS $e^+e^-\gamma$ corrections are realized in Ref.~\cite{Jadach:1992nk} 
  by second order structure function with the running coupling.
}
and Singlet structure functions. 
Contrary to Ref.~\cite{Arbuzov:1995qd}, 
results are provided also for the asymmetric acceptances.

The approach of \cite{Jadach:1996ca} was based on the 
extension of the YFS~\cite{Yennie:1961ad}
scheme of the soft photon resummation to the case of soft $e^+e^-$ pair emission,
with relevant real and virtual soft ingredients calculated in \cite{Jadach:1993wk}
(omitting up-down interference, multi-peripheral graphs {\em etc.}). 
The calculation is implemented in the unpublished BHLUMI v.\ 2.30 MC code. 
The accuracy of results was estimated  to be 0.02\% 
for the asymmetric angular acceptance, 
i.e. $3.3^\circ - 6.3^\circ$ and $2.7^\circ - 7.0^\circ$,  
with the energy cut $1-s'/s <z_{cut} =0.5$. 
Ref.~\cite{Montagna:1998vb} has concluded that
this precision is even better, $6\times 10^{-5}$for $z_{cut} \leq 0.5$, while
for hard emission, $z_{\rm cut}> 0.5$,
with significant multi-peripheral component,
the precision deteriorates to 0.01\%.

{\em What should be done in order to consolidate
the above, mostly LEP era, calculations of the fermion pair contribution
and to reach even better precision level needed for FCC-ee?}
%\begin{enumerate}
%\item

As in ref.~\cite{Montagna:1998vb}, for the additional real 
$e^+e^-$ pair radiation the complete matrix element should be used,
because non-bremsstrahlung-type graphs can 
contribute as much as 0.01\% for the cut-off $z_{cut}\sim 0.7$. 
There is a number MC generators for the $e^+e^-\to 4f$ process, 
developed for the LEP2 physics to be exploited for that purpose%
\footnote{
  One needs to be sure that the collinear configurations of
  outgoing four electrons are covered, 
  for example like it is done in KoralW \cite{Jadach:1998gi}
  which in addition, in its latest version 1.53 \cite{Jadach:2002hh}, 
  accounts for  photonic radiation to t-channel exchanges as well.}.
%%%%%%%%%%%%%%%%%%%%%%%%%%%%%%%%%
%\item

In order to improve on 0.005\% uncertainty of Ref.~\cite{Arbuzov:1995qd},
due to the emission of the $\mu^+\mu^-$, $\tau^+\tau^-$, and quark pairs,
one may use LO  calculation of ref.~\cite{Jadach:1992nk},
incorporating lepton pair contributions by means of
the modification of the running coupling.
A naive rescaling of the electron logarithm
(due to the mass of the muon)
%(due to wider angular range of the FCC-ee luminometer??)
gives  $\ln\frac{|t|}{m_e^2}=17.5$ and $\ln\frac{|t|}{m_\mu^2}=6.9$
i.e.\ for muon pairs we find a suppression factor of 
$\ln^2\frac{|t|}{m_\mu^2}/\ln^2\frac{|t|}{m_e^2}=0.4^2=0.16$ 
relative to the electron pair.
Rescaling additional $e^+e^-$ pair contribution of 0.05\% one obtains
an estimate of the muon pair contribution of 0.008\%. 
\footnote{
This is less optimistic than the estimate in Ref.~\cite{Arbuzov:1995qd}.
}.
For the tau lepton logarithm $\ln\frac{|t|}{m_\tau^2}=1.2$ 
we obtain $\ln^2\frac{|t|}{m_\tau^2}/\ln^2\frac{|t|}{m_e^2}=0.07^2=0.005$ 
suppression factor relative to the electron pair,
hence this contribution can be neglected.
Adding $\mu^+\mu^-$ pairs to the BHLUMI v.\ 2.30 code
of Ref.~\cite{Jadach:1993wk} would be straightforward. 
Also in the  approach of ref.~\cite{Montagna:1998vb,Montagna:1999eu} 
this should be possible%
\footnote{
  The other option is to use the above described 
  general purpose LEP2 $4f$ codes, including also
  the discussed earlier corresponding virtual corrections.
  }.
The contribution of light quark pairs ($\pi$ pairs etc.) 
can be roughly estimated using quantity 
R$_{had} = \sigma_{had}/\sigma_{\mu}\simeq 3$
for the effective hadronic production threshold of the order of 1 GeV. 
One obtains 
R$_{had}\ln^2\frac{|t|}{0.5^2 GeV^2}/\ln^2\frac{|t|}{m_\mu^2}= 0.9$, 
i.e.\ this contribution is of the size of the muon pair contribution,
that is  of the order of 0.008\%. 
\footnote{
This is less optimistic than the estimate in Ref.~\cite{Arbuzov:1995qd}.
Adding in quadrature errors due to muon and light quark pairs one obtains 
0.011\% rather than 0.006\% of Ref.~\cite{Arbuzov:1995qd}. 
0.011\% is consistent with the estimate of Ref.~\cite{Montagna:1998vb}.
}.
%For the moment let us treat light quark pairs 
%contribution as a component of the error being 0.008\%.
%%%%%%%%%%%%%%%%%%%%%%%%%%%%%%%%%%%%%%%%
%\item

The third group of corrections are the higher order terms. 
The emission of two (or more) electron pairs is suppressed by another 
factor $(\frac{\alpha} {\pi} \ln\frac{|t|}{m_e^2})^2 \sim 10^{-3}$ 
and is negligible. 
The additional $e^+e^- +n\gamma$ correction is non-negligible. 
Its evaluation was based 
either on LO structure functions (\cite{Arbuzov:1995qd} (Table~1), 
\cite{Montagna:1998vb} (Fig.~8), \cite{Jadach:1992nk}) 
or on the YFS~\cite{Yennie:1961ad}
soft approximation \cite{Jadach:1996ca} (Fig.~4),
resulting in quite different results
and their comparison is rather inconclusive.
They are at most of the order of 0.5 to 0.75 of the additional $e^+e^-$ correction
(without $\gamma$). 
The remaining non-leading, non-soft additional $e^+e^-+n\gamma$ corrections 
are suppressed by another $1/ \ln\frac{|t|}{m_e^2} \sim 0.06$ 
and should be negligible ($\sim 0.003\%$). 
It would be also possible 
to calculate the additional $e^+e^- +\gamma$ real emission in a way similar to 
the existing code for LEP2 physics, \cite{Denner:2002cg}
%\end{enumerate}

The above improvements can be implemented either 
directly in the upgraded \bhlumi\
or using a separate calculation, 
such as \bhlumi~2.30~\cite{Jadach:1996ca} code,
or external MC programs like 
these of Refs.~\cite{Montagna:1998vb,Montagna:1999eu}.
To summarize, the proposed future error budget is the following:
(1) The contribution of light quark pairs must be calculated with the accuracy 
of 25\%, i.e.\ 0.0027\%. 
(2) The contribution of the muon pairs will be known to 10\%,
i.e.\ to 0.0008\%. (3) The non-leading, non-soft additional 
$e^+e^-+n\gamma$ corrections
will be treated as an error of 0.003\%. 
Adding (1)--(3) in quadrature we obtain 0.004\%.
Applying safety factor of 1.25 we end up with 0.005\% possible pair production 
uncertainty forecast for the FCC-ee, quoted in Table~\ref{tab:lep2fcc}.

\subsection{{\em Z} exchange and {\em s}-channel photon exchange}
\label{bhlumizgamma}
%%%%%%%%%%%%%%%%%%%%%%%%%%%%%%%%%%%%%%%%%%%%%%%%%%%%%%%%%%%%%

In the Bhabha scattering process, in addition to $\gamma$ exchange in the
$t$ channel $\gamma_t$, there are also contributions from $\gamma$ exchange in the $s$ 
channel $\gamma_s$ and $Z$ exchange in both $t$ and $s$ channels, $Z_t$ and $Z_s$. 
In fact, they all should be added at the amplitude level (Feynman diagram)
and then squared to obtain the differential cross section for the Bhabha process,
giving rise to several interference contributions.
Numerically the most important for the low angle Bhabha (LABH) luminometry, 
apart from the pure $t$-channel $\gamma$ exchange, $\gamma_t\otimes \gamma_t$,
are interferences of other contributions with the $\gamma_t$ amplitude,
due to the enhancement factor $\sim s/|t|$. 
Among these, near the $Z$ peak, the most sizable 
is the interference $\gamma_t \otimes Z_s$,
because of the resonant  enhancement.
In the context of LEP luminometry it was studied in detail in
ref.~\cite{Jadach:1995hv} for two types of detectors: 
SICAL with an angular coverage of $\sim 1.5^{\circ}$--$3^{\circ}$ 
and LCAL with an angular coverage of $\sim 3^{\circ}$--$6^{\circ}$. 
Based on this work, the $\gamma_t \otimes Z_s$ contribution 
was implemented in \bhlumi\ 4.02 and its theoretical precision for the LEP
luminosity measurement was assessed.
We are going to exploit these results and estimate
theoretical errors of all other contributions beyond the dominant
$\gamma_t\otimes \gamma_t$.
Since the angular coverage of the planned FCC-ee 
luminometer~\cite{MogensDam:2018} 
is close to the LCAL one,
we shall use the results of ref.~\cite{Jadach:1995hv} 
obtained for this type of the detector.

The Born-level $\gamma_t \otimes Z_s$ contribution is
up to $\sim 1\%$ and changes from being positive 
below the $Z$ peak to negative above, 
reaching the maximal absolute value at 
about $\pm 1\,$GeV from the peak. Radiative corrections, dominated by QED,
are sizable, up to $\sim 0.5\%$ 
(up to $\sim 50\%$ of the Born-level contribution)
and change in the opposite way, 
i.e.\ from negative to positive values
when going from below to above the $Z$ peak. 
\bhlumi\ includes the QED corrections and running-coupling effects 
for this contribution within the ${\cal O}(\alpha)$ YFS exclusive exponentiation.
The theoretical uncertainty for this calculation was estimated at $0.090\%$ for LCAL
and is used as an initial estimate of the theoretical error for
the FCC-ee luminometry concerning the $\gamma_t \otimes Z_s$ contribution
in Table~\ref{tab:lep2fcc}.

The other contributions will be estimated by means
of relating them to the $\gamma_t \otimes Z_s$ or $\gamma_t \otimes \gamma_t$,
using rescaling factors,
$|t|/s \approx 1.3\times 10^{-3}$ and
$\tilde{\gamma}_Z = \Gamma_Z/M_Z \approx 2.7\times 10^{-2}$.

The next most sizable contribution comes from the interference
$\gamma_t\otimes\gamma_s$.
At the Born level, near the $Z$ peak, it is smaller than the 
$\gamma_t \otimes Z_s$ contribution by the factor%
\footnote{The factor of $4$ comes from the ratio of 
   the corresponding coupling constants.}
$\sim 4\,\tilde{\gamma}_Z \approx 0.1$.
Taking $\sim 1\%$ for the Born-level $\gamma_t \otimes Z_s$,
we get $\sim 0.1\%$ for $\gamma_t\otimes\gamma_s$.
It is included in \bhlumi, 
so we need to estimate the missing radiative corrections.
Since this is smooth near the $Z$ peak,
the photonic QED corrections should stay within $10\%$,
for not too tight cuts on radiative photons. 
The resulting estimate of
the theoretical precision of $\gamma_t\otimes\gamma_s$ contribution
in \bhlumi\ for the FCC-ee luminometry is $\sim 0.01\%$.

The resonant pure $s$-channel $Z$ contribution, $Z_s \otimes Z_s$,
at the Born level, is multiplied with respect to the 
$\gamma_t \otimes Z_s$ term by the factor 
$\sim |t|/s\times 1/(4\,\tilde{\gamma}_Z) \approx 1.3\times 10^{-2}$,
thus its size is $\sim 0.01\%$.
It is omitted in the current version of \bhlumi, 
hence it enters into theoretical error as a whole.
However, it can be included rather easily,
such that only the missing radiative corrections will matter.
Due to the $Z$-resonance effect, they
can reach even $\sim 50\%$ of the Born-level contribution,
hence the corresponding theoretical error would be $\sim 0.005\%$.

The $t$-channel interference $\gamma_t \otimes Z_t$ 
we estimate multiplying the $\gamma_t \otimes Z_s$ contribution by
the $\sim |t|/s \times \tilde{\gamma}_Z \approx 3.5\times 10^{-5}$ factor.
It can be easily implemented in \bhlumi,
with the theoretical error due
to the missing photonic corrections being below $10^{-5}$.

The pure $s$-channel $\gamma_s\otimes\gamma_s$ contribution 
is much smaller in the $Z$-peak region than the resonant $Z$ exchange.
It is suppressed by the factor
$\sim (4\,\tilde{\gamma}_Z)^2 \approx 0.01$ with respect to $Z_s \otimes Z_s$
(which is worth $\sim 0.01\%$),
so is of the order of $10^{-6}$.

Finally, the $Z_t \otimes Z_t$ contribution is smaller than the dominant 
$\gamma_t \otimes \gamma_t$ one by the factor $\sim (|t|/s/4)^2 < 10^{-6}$,
thus it is completely negligible.

Adding the above theoretical errors in the quadrature,
we obtain the total uncertainty (contributions omitted in \bhlumi)
due to the $Z$ exchanges and $\gamma_s$ exchange 
for the FCC-ee luminometer near the $Z$ peak at the level of $0.090\%$,
quoted as present state of the art in Table~\ref{tab:lep2fcc}.

The above uncertainty is completely dominated by the uncertainty of the
$\gamma_t \otimes Z_s$ contribution which comes from a rather conservative
estimate in ref.~\cite{Jadach:1995hv} 
based on comparisons of \bhlumi\ with the MC generator 
{\tt BABAMC}\cite{Berends:1987jm} and the semi-analytic program
{\tt ALIBABA} \cite{Beenakker:1990mb,Beenakker:1990es},
the latter including higher-order leading-log QED effects.
Later on, the new MC event generator {\tt BHWIDE} \cite{Jadach:1995nk} 
was developed for the wide-angle Bhabha scattering
including all Born-level contributions for Bhabha process and 
${\cal O}(\alpha)$ YFS exponentiated EW radiative corrections.
The comparison of \bhlumi\ with {\tt BHWIDE} for FCC-ee luminometer
would help to reduce all the above theoretical errors.
In principle, the Born-level but also ${\cal O}(\alpha)$ QED
matrix elements of {\tt BHWIDE} could be implemented in \bhlumi.
This would reduce the theoretical error 
for the above group of contributions below $0.01\%$,
as indicated in Table~\ref{tab:lep2fcc}.
What we can do right now is to examine in a more detail
the $\gamma_t \otimes Z_s$ contribution using {\tt BHWIDE},
in order to get better idea about its future uncertainty.
The main advantage of {\tt BHWIDE} is that its matrix element includes
complete ${\cal O}(\alpha)$ corrections to $\gamma_s$, $Z_s$ and $Z_t$
exchanges, while \bhlumi\ includes only part of ${\cal O}(\alpha)$
corrections due to soft photon resummation.

\subsection{Study of {\em Z} and {\em s}-channel $\gamma$ exchanges 
            using {\tt BHWIDE}}
\label{bhlumizgamma2}
%%%%%%%%%%%%%%%%%%%%%%%%%%%%%%%%%%%%%%%%%%%%%%%%%%%%%%%%%%%%%

%%%%%%%%%%%%%%%%%%%%%%%%%%%%%%%%%%%%%%%%%%%%%%%%%%%%%%%%%%%%%%%%%%%%%%%%%%%%%%%
%%%%%%%%%%%%%%%%%%%%%%%%%%%%%%%%%%%%%%%%%%%%%%%%%%%%%%%%%%%%%%%%%%%%%%%%%%%%%%%
\begin{table}[!ht]
\centering
\begin{tabular}{|c|c|c|c|c|}
\hline %%%%%%%%%%%%%%%%%%%%%%%%%%%%%%%%%%%%%%%%%%%%%%%%%%%%%%%%%%%
$E_{\rm CM}$ [GeV]
    & $\Delta_{\rm tot}$ [\%]
         & $\delta_{{\cal O}(\alpha)}^{\rm QED}$ [\%]
              & $\delta_{\rm h.o.}^{\rm QED}$ [\%]
                   & $\delta_{\rm tot}^{\rm weak}$ [\%]
\\  \hline %%%%%%%%%%%%%%%%%%%%%%%%%%%%%%%%%%%%%%%%%%%%%%%%%%%%%%%%%%
%\\ %%%%%%%%%%%%%%%%%%%%%%%%%%%%%%%%%%%%%%%%%%%%%%%%%%%%%%%%%%%%%%%%%%
$90.1876$       
  & $+0.642\,(12)$ 
    & $-0.152\,(59)$
      & $+0.034\,(38)$
        & $-0.005\,(12)$
\\ %%%%%%%%%%%%%%%%%%%%%%%%%%%%%%%%%%%%%%%%%%%%%%%%%%%%%%%%%%%%%%%%%%
%\\ %%%%%%%%%%%%%%%%%%%%%%%%%%%%%%%%%%%%%%%%%%%%%%%%%%%%%%%%%%%%%%%%%%
$91.1876$       
  & $+0.041\,(11)$ 
    & $+0.148\,(59)$
      & $-0.035\,(38)$
        & $+0.009\,(12)$
\\ %%%%%%%%%%%%%%%%%%%%%%%%%%%%%%%%%%%%%%%%%%%%%%%%%%%%%%%%%%%%%%%%%%
%\\ %%%%%%%%%%%%%%%%%%%%%%%%%%%%%%%%%%%%%%%%%%%%%%%%%%%%%%%%%%%%%%%%%%
$92.1876$       
  & $-0.719\,(13)$ 
    & $+0.348\,(59)$
      & $-0.081\,(38)$
        & $+0.039\,(13)$
\\ \hline %%%%%%%%%%%%%%%%%%%%%%%%%%%%%%%%%%%%%%%%%%%%%%%%%%%%%%%%%%%
\end{tabular}
%30Aug \setcounter{table}{0}
\caption{\sf
Results from \bhwide\ for the $Z$ and $\gamma_s$ exchanges
contribution to the FCC-ee luminosity with respect to the
$\gamma_t\otimes\gamma_t$ process for the calorimetric
LCAL-type detector \cite{Jadach:1991cg} 
with the symmetric angular range $64$--$86\,$mrad; 
no acoplanarity cut was applied.
MC errors are marked in brackets.
}
\label{tab:Zsgam}
\end{table}
%%%%%%%%%%%%%%%%%%%%%%%%%%%%%%%%%%%%%%%%%%%%%%%%%%%%%%%%%%%%%%%%%%%%%%%%%%%%%%%

We are going to present
numerical results obtained with \bhwide\ for the calorimetric LCAL-type
detector, as described in ref.~\cite{Jadach:1991cg},
for the symmetric angular range $64$--$86\,$mrad without
any cut on acoplanarity (i.e.\ the number of azimuthal
sectors in LCAL was set to $1$). For the $Z$-boson mass and width we
used the current PDG values: $M_Z=91.1786\,$GeV
and $\Gamma_Z=2.4952\,$GeV. 
The weak corrections, i.e.\ the non-QED electroweak (EW) ones, 
were calculated with the help of the {\tt ALIBABA} 
EW library~\cite{Beenakker:1990mb,Beenakker:1990es}.
The results, shown in Table~\ref{tab:Zsgam},
were obtained for three values of the centre-of-mass (CM)
energy: $E_{\rm CM} = M_Z,\, M_Z\pm 1\,$GeV. 
The last two values were chosen because for these energies 
the $Z$-contribution is (close to) the largest -- with the opposite sign, 
see e.g.\ ref.~\cite{Jadach:1995hv}.

The numbers shown in the second column of Table~\ref{tab:Zsgam} represent
the total relative contribution of the $Z$ and $\gamma_s$ exchanges,
$\Delta_{\rm tot}= |\gamma_t +\gamma_s +Z_s +Z_t|^2/|\gamma_t|^2-1$,
as predicted by \bhwide, 
that is for Born + ${\cal O}(\alpha)$ YFS exponentiated matrix elements,
including ${\cal O}(\alpha)$ EW corrections.
As one can see, this contribution is positive below the $Z$-peak
with the size up to $\sim 0.64\%$, gets close to zero near the $Z$-peak
and changes the sign above the $Z$-peak with the size up to
$\sim -0.72\%$. 
This agrees with our rough estimate given in the previous 
subsection that the Born-level contribution is up to about $1\%$.
 
These effects are in general consistent with the results of 
ref.~\cite{Jadach:1995hv}, although they are slightly smaller. 
There are two main reasons for this: 
(1) the polar angles of the LCAL detector are a bit smaller here 
than in ref.~\cite{Jadach:1995hv} and 
(2) here we used the calorimetric acceptance, while the results
in Tables 1 and 2 of ref.~\cite{Jadach:1995hv} 
were obtained for the non-calorimetric acceptance. 

In the next three columns we present various interesting
components of the radiative corrections in $\Delta_{\rm tot}$.
The fixed-order (without exponentiation) ${\cal O}(\alpha)$ QED
corrections, shown in the third column, are sizable -- from
about $-0.15\%$ below $M_Z$ to about $+0.35\%$ above it, and
they have the opposite sign to the Born-level contribution. 
This also agrees with our estimate given above that the QED
correction can reach up to a half of the size of the Born-level effect. 

In the fourth column we show the higher-order QED corrections,
i.e.\ the ones beyond the ${\cal O}(\alpha)$ QED fixed-order,
which result from the YFS exponentiation. 
They also change their sign near the $Z$-peak, but in the opposite
way to the ${\cal O}(\alpha)$ corrections, and their size is about
a quarter of the latter.
Based on the size of these corrections one can estimate the
higher-order QED effects missing in \bhwide. 
Since near the $Z$-peak the dominant are soft-photon corrections
which are treated by the YFS exponentiation very accurately,
we may expect that those missing effects are much smaller
than the ones in the fourth column of Table~\ref{tab:Zsgam}.
To estimate their size we can use the factor 
$\gamma=\frac{\alpha}{\pi}\ln\frac{|\bar{t}|}{m_e^2}=0.042$
of Section~\ref{sec:lumi4} and the `safety' factor 2 
of ref.~\cite{Jadach:1995hv}, and apply them to the largest
h.o.\ correction in Table~\ref{tab:Zsgam}, i.e.
$0.081\%\times \gamma \times 2 \simeq 0.007\%$.

Note that precision of the MC results in Table~\ref{tab:Zsgam}
is limited by the statistical error of order $0.01-0.06\%$, 
because MC sample from \bhwide\ in the current
version of the program is limited due to the size limit
of the Fortran integer numbers.
Nevertheless, these results provide useful estimates
on the size of the higher order effects.

In the last column of Table~\ref{tab:Zsgam} the pure EW corrections,
i.e.\ the EW corrections minus the QED ones are shown --
implemented within the ${\cal O}(\alpha)$ YFS exponentiation scheme.
They are at the level of about $0.01\%$ below and at $M_Z$,
while above $M_Z$ they increase up to $\sim 0.04\%$.

To estimate the size of the missing higher-order weak corrections
in \bhwide, we can apply the same factor as for the QED corrections
to get the value of $\sim 0.003\%$. 

Altogether, we can estimate the physical precision of
the $Z$ and $\gamma_s$-exchanges contribution 
to FCC-ee luminometry in \bhwide\ by adding linearly 
(to be conservative) 
the above two numbers on the missing effects to get $\sim 0.01\%$.
Therefore, if the predictions of \bhlumi\ for the luminosity
measurement at FCC-ee are combined with the ones from \bhwide\ for
this contribution, then the error in the line (e) of
Table~\ref{tab:lep2fcc} could be reduced to $0.01\%$.
Of course, this result requires more dedicated numerical tests
and cross-checks with independent calculations.

In ref.~\cite{Battaglia:2001dg} it was shown that for
$\sqrt{s} \gg  M_Z$ all the above contributions are below $0.01\%$ 
-- they then can be neglected in the FCC-ee luminometry  
at energies above the $Z$ peak.

The best method to reduce the uncertainty of the above contributions
practically to zero would be to include these $Z$ and $\gamma_s$ exchanges
within the CEEX matrix element at \order{\alpha^1} in \bhlumi.
Most likely, it would be enough to add the EW corrections
to the LABH process in the form of effective couplings in the Born amplitudes.
On the other hand, the new \bhlumi\ with such a CEEX matrix element
would serve as a starting point for a much better
wide-angle Bhabha MC generator, similarly as \bhlumi~v. 4.04 has served
as a starting point for {\tt BHWIDE} \cite{Jadach:1995nk}.

\section{Technical precision}
%%%%%%%%%%%%%%%%%%%%%%%%%%%%%%%%%%%%%%%%%%%%%%%%%%%%%%%%%%%%%
\label{sec:tech-prec}
The question of the technical precision is quite nontrivial
and difficult.
The evaluation of the technical precision of \bhlumi\ v.4.04
with YFS soft-photon resummation and complete \order{\alpha^1}
relies on two pillars: the comparison with semi-analytic
calculations done in ref.~\cite{Jadach:1996bx} and comparisons
with two hybrid MC programs {\tt LUMLOG+OLDBIS} and {\tt SABSPV},
reported in ref.~\cite{Jadach:1996gu}.
This precision was established to be 0.027\%
(together with missing photonic corrections).
Note that this was not an ideal solution, because the above
two hybrid MCs did not feature complete soft photon 
resummation and disagreed with \bhlumi\ by more than $0.17\%$ 
for sharp cut-offs on the total photon energy.

In fact, after the LEP era, another MC program \babayaga%
~\cite{CarloniCalame:2000pz,CarloniCalame:2001ny,Balossini:2006wc},
with soft-photon resummation has been developed 
using a parton shower (PS) technique, 
and in principle could be used for better validation
of the technical precision of both \bhlumi\ and \babayaga.
In fact, such a comparison with {\tt BHWIDE} MC~\cite{Jadach:1995nk}
was done for $s^{1/2} \leq 10\,$GeV
and the $0.1\%$ agreement was found. It is quite likely
that such an agreement persists near $s^{1/2}=M_Z$.

Let us note in passing that the inclusion of the complete \order{\alpha^1}
into \babayaga\ was done before three technologies of matching
fixed-order NLO calculations with a parton shower (PS) algorithm 
were unambiguously established:
{\tt MC@NLO}~\cite{Frixione:2002ik},
{\tt POWHEG}~\cite{Nason:2004rx} and {\tt KrkNLO}~\cite{Jadach:2015mza}.
The algorithm of NLO matching in \babayaga\ 
is quite similar to that of {\tt KrkNLO}%
\footnote{Single MC weight is introducing NLO correction in both methods,
but in {\tt KrkNLO} it sums over real photons,
while in \babayaga\ it takes product over them. 
However, it is the same when truncated to \order{\alpha^1}.
We are grateful to authors of \babayaga\ for clarification on this point.}

Ideally, in the future validation of the upgraded \bhlumi,
in order to get its technical precision
at the level $10^{-5}$ for the total cross section and $10^{-4}$
for single differential distributions, one would need to compare
it with another MC program developed independently,
which properly implements the soft-photon resummation,
LO corrections up to \order{\alpha^3 L_e^3}, and
the second-order corrections with the complete \order{\alpha^2 L_e}.

In principle, an extension of a program like \babayaga\
to the level of NNLO for the hard process,
keeping the correct soft-photon resummation,
would be the best partner for the upgraded \bhlumi\ to establish the
technical precision of both programs at the $10^{-5}$
precision level%
\footnote{ The upgrade of the \bhlumi\ distributions will be
  relatively straightforward because its multi-photon 
  phase space is exact~\cite{Jadach:1999vf} for any number of photons.}.
In the meantime, the comparison between the upgraded
\bhlumi\ with EEX and CEEX matrix elements would also offer
a very good test of its technical precision,
since the basic multi-photon phase space integration module of \bhlumi\
was already well tested in ref.~\cite{Jadach:1996bx}
and such a test can be repeated at an even higher-precision level.

\section{Summary}
\label{sec:lumi5}
%%%%%%%%%%%%%%%%%%%%%%%%%%%%
Summarizing, we conclude that an upgraded new version of \bhlumi\ 
with the error budget of 0.01\% shown in Table~\ref{tab:lep2fcc}
is perfectly feasible.
With appropriate resources, such a version of \bhlumi\
with the \order{\alpha^2} CEEX matrix element and
with the precision tag of $0.01\%$, needed for the FCC-ee physics,
could be realized.
A new study of the $Z$ and $s$-channel $\gamma$ exchanges using \bhwide\ MC
was instrumental in the above analysis.
Keeping in mind that the best experimental error
of luminosity measurement achieved at LEP was 0.034\% \cite{Abbiendi:1999zx},
it would be interesting to study whether the systematic error
of the designed FCC-ee luminosity detector \cite{MogensDam:2018}
can match the above anticipated theory precision.

\vspace{5mm}
\noindent
{\bf\large Acknowledgments}\\
We would like to thank F. Jegerlehner for exhaustive discussion on
the issues related to the vacuum polarization and to the Pavia group
members, F.~Picinnini, C.~C.~Calame, G.~Montagna and O.~Nicrosini,
for the explanations concerning the \babayaga\ MC program.

%\newpage
%\bibliography{lumi}{}
%\bibliographystyle{utphys_spires}
%\bibliographystyle{utphys_spires_tit}
%\bibliographystyle{jhep}
%\bibliographystyle{plain}

\providecommand{\href}[2]{#2}
%%%%%%%%%%%%%%%%%%%%%%%%%%%%%%%%%%%%%%%%%%%%%%%%%%%%%%%%%%%%%%%%%%%
\end{document}